Parity-dependent shot noise and spin-flip relaxation process in a hybrid superconductor-nanowire quantum dot

Keiko Takase[1*], Yasuhiro Utsumi[2], Yamato Ashikawa[1,3], Guoqiang Zhang[1], Kouta Tateno[1], Yuma Okazaki[1] and Satoshi Sasaki[1]

[1] *NTT Basic Research Laboratories, NTT Corporation,*
*3-1 Morinosato-Wakamiya, Atsugi, Kanagawa 243-0198, Japan*
[2] *Department of Physics Engineering, Faculty of Engineering, Mie University, Tsu, Mie, 514-8507, Japan*
[3] *Department of Physics, Tohoku University, 6-3 Aramaki Aza Aoba, Aoba-ku, Sendai 980-8578, Japan*

**Abstract**

We report shot noise measurements for a quantum dot formed in an InAs nanowire suspended between superconducting electrodes. We find a clear alternation for the shot noise value in the Coulomb blockade regime between even and odd electron occupation in the dot, indicating that super-Poissonian (Poissonian) shot noise with the Fano factor reaching around 2 (1) occurs for even (odd) parity. With increasing magnetic field, the parity effect disappears and all the regimes show the Fano factor of around 1. The whole observation in our experiments quantitatively agrees with simulation obtained from full-counting statistics of cotunneling including spin-flip relaxation process, which corresponds to modelling electron motion in a quantum dot with strong spin-orbit interaction.

**Texts**

Shot noise is a time-dependent current fluctuation that reflects the discreteness of charge carriers [1], thereby being used to reveal electron statistics and the effective charge of quasi-particles in mesoscopic systems in a variety of materials [2,3], such as the fractional quantum Hall effect in GaAs [4] and the Andreev bound states formed in a one-dimensional InAs nanowire channel with superconductor [5]. For a zero-dimensional quantum dot, shot noise spectroscopy has been useful to elucidate microscopic electron motion in the Coulomb blockade regime, where the first-order tunneling current cannot flow.

One of the most intriguing phenomena in the Coulomb-blockade regime is elastic and inelastic cotunneling, which is a higher-order tunneling process involved with ground state and excited states in the dot. So far, the cotunneling accompanied by dynamical charge fluctuation has been intensively investigated. Theoretically, many groups predict the presence of the super-Poissonian shot noise [6–14] associated with inelastic cotunneling, which is found to be consistent with many experiments done for a quantum dot fabricated from diverse materials such as GaAs [15,16], Si [17], and CNT [18]. On the other hand, cotunneling inducing super-Poissonian noise has not been investigated for the system with strong spin-orbit interaction, whereas the demonstration of charge-fluctuation sensing and the detection of higher-order processes are increasingly important in association with not only a spin-orbit interaction based qubit [19] but also the recently emerged fields of Majorana physics [20,21] and topological devices [22,23].

Here, we report shot-noise spectroscopy on a suspended InAs nanowire quantum dot bridging two superconducting electrodes and investigate the quasi-particle cotunneling regime. We found that the Fano factor, which is proportional to the ratio of shot noise to current, alters between even and odd electron occupation in a quantum dot, indicating that super-Poissonian (Poissonian) shot noise with Fano factor larger than (equal to) 1 being associated with inelastic (elastic) cotunneling transport occurs for even (odd) occupation. With increasing magnetic field, the parity effect and the super-Poissonian shot noise disappear. Our experiments are quantitatively explained by simulation accompanied by spin-flip relaxation, which is not included in previous theories [6–14]. Our experiments and simulation pave the way to revealing microscopic electron motions for quantum devices employing the spin-orbit interaction.

The sample used here is a suspended InAs nanowire attached to two superconducting electrodes. The InAs nanowires were grown by the vapor-liquid-solid method [24]. The left panel in Fig. 1 shows a SEM image of the top view of the sample. A quantum dot is formed in the area confined between the two superconducting electrodes (Ti/Al), which are

100 nm apart. The right panel shows a side view of the sample. The nanowires were transferred onto the pre-patterned wafer, and then HSQ and PMMA, both of which are known to give less damage to samples [25–27], were spin coated in sequence. We then fabricated Ti/Al (5 nm/45 nm) source-drain electrodes, followed by the removal of the HSQ layer outside the electrode regions. This left an air gap of about 30 nm between the nanowire and the gate electrode to minimize the environmental disorder. Fabrication methods used for similar devices and basic transport properties on our nanowires have been reported in details elsewhere [28–32].

The measurements were carried out using a setup that enables low-frequency lock-in measurements and shot noise measurements to be performed simultaneously at 50 mK, which is schematically shown in Fig. 1. The frequency used for the shot noise measurements was around 3 MHz, which was high enough to allow us to disregard the contribution from $1/f$ noise. To deduce the shot noise, we followed the manner similar to that described in Refs. [16] and [33]. We tuned the gate voltage to the regime where the dot lead coupling is relatively strong to observe the super-Poissonian shot noise in the cotunneling regime.

We also examine full-counting statistics for a quantum dot with superconducting electrodes. The basic calculation scheme follows the procedure in Refs. [6] and [7]. In contrast, what is very different from previous theories reporting super-Poissonian shot noise [6–11] is that we newly include the spin-flip relaxation process in our model, which is expected to appear in devices with strong spin-orbit interaction. While this calculation method is less systematic than the real-time diagrammatic theory [8–10,12,34], our method is simpler and is able to demonstrate the presence of the super-Poissonian noise in the Coulomb blockade regime, as is consistent with the previous theories [6–14]. The details for the calculation is written in the Supplemental Material.

Figure 2 shows the differential conductance plotted as a function of gate voltage ($V_g$) and source drain bias ($V_{sd}$), where a familiar pattern of sequential Coulomb diamonds is observed. We find that the sizes of neighboring Coulomb diamonds, which reflect the energy required to add one electron to a quantum dot, are different [Fig. 2(a)]; generally the Coulomb diamonds observed for even occupation are larger than those for odd occupation, since an extra energy is required for an electron to enter the quantum dot with the highest spin degenerate energy level already occupied [35].

In the Coulomb blockade regime, we observed two peaks that were symmetric with respect to $V_{sd} = 0$, as is more clearly seen in the cut line in Fig. 2(b). The energy spacing between the peaks was 360 mV, corresponding to twice the superconducting gap of Ti/Al electrodes $2\Delta_s \sim 180$ meV. The peaks with an energy spacing of $2\Delta_s$ that have been observed for SQS junctions are known to occur as a result of elastic cotunneling by which a quasi-particle from the superconducting electrode tunnels via a higher-order process [36], as is shown schematically in Fig. 2(c). When the magnetic field ($B$) is increased, the cotunneling peaks weaken and entirely disappear under 0.2 T [Fig. 2(d)], which is larger than the critical magnetic field of Ti/Al. This indicates that the origin of these peaks is certainly associated with the superconducting electrodes.

Figure 3 shows the results of shot noise measurements (bottom panel) in the form of Fano factor ($F$) mapping, together with those of time-averaged low-frequency lock-in transport (top panel), which were measured simultaneously for almost the same $V_g$ range as in Fig. 2. Here $F$ is defined as $F = S/2eI$, where $S$ is the Fourier power spectral density of the time-domain current fluctuation, $e$ is the electron charge, and $I$ is the time-averaged current. By comparing the top and bottom panels, we can clearly see that, outside the Coulomb diamonds, the Fano factor is suppressed to around $F \sim 0.5$, being sub-Poissonian noise ($F < 1$). The feature of $F < 1$ outside the Coulomb diamonds has been observed in several experiments [15,16], the mechanism of which is known to be the first-order sequential tunneling.

More importantly, the Fano factor in the Coulomb diamonds shows a parity-dependence, $F \sim 2$ for the even occupation and $F \sim 1$ for the odd occupation, which indicates that super-Poissonian (Poissonian) statistics occurs for the even (odd) parity. The region where $F$ is large can be found in $V_{sd} > 2|\Delta_s|$, and in some cases, in the area where $V_{sd}$ is larger than the dashed lines, which connect with the Coulomb peaks (dash-dotted lines) outside the Coulomb diamonds. These Coulomb peaks are reported to be associated with the excited states [35], suggesting that the occurrence of the super-Poissonian shot noise is strongly related with the excited states.

To understand the mechanism systematically, we have performed shot noise measurements from zero to finite magnetic fields, and also compared the results with simulation of Fano factor. Figure 4(a) shows shot noise spectroscopy obtained at zero and 0.9 T (>critical field of the superconducting gap) as well as differential conductance that were simultaneously measured. The panels of the differential conductance show that, with increasing $B$, the sizes of the Coulomb diamonds become smaller and the size difference depending on the parity of the electron occupation also becomes smaller. This is reasonably understood by considering energy diagram of the dot [Fig. 4(c)]. Since the sizes of the Coulomb diamonds are determined by charging energy $U$ and level spacing $\Delta E$, the sizes are given at zero field by $W = U + \Delta E$ ($W = U$) for the even (odd) occupation apart from superconducting gap $2\Delta_s$. On the other hand, with increasing $B$, the Zeeman gap opens and the diamond sizes are $W = U + \Delta E - \Delta Z$ ($U + \Delta Z$) for even (odd) parity ($\Delta Z$: Zeeman energy), leading to smaller difference in size between even and odd occupations. Turning to shot noise data, the parity effect, i.e. alternation of super-Poissonian ($F \sim 2$) and Poissonian ($F \sim 1$) that are observed at zero field disappears at $B = 0.9$ T. This results in negligible even-odd difference for $F$, giving rise to $F \sim 1.0$ to 1.1 for all the diamonds.

We next consider the underlying mechanism. Because $U < \Delta E$ is estimated from the sizes of the neighboring Coulomb diamonds, the schematic diagram that illustrates the Coulomb diamonds for our experiments are shown in Fig. 4(c). Based on a previous experiment done for a GaAs quantum dot, which shows super-Poissonian shot noise in $\Delta E < eV_{sd} < W$ [16], our measurement conditions lead to the expectation that the super-Poissonian shot noise can be observed in the regime painted in Fig. 4(c), giving rise to the parity effect at zero field. Actually, our experimental results at zero field show the parity effect, but observation at finite fields are different from the simple qualitative expectation; both for even and odd parity, the super-Poissonian shot noise reaching $F \sim 2$ is not observed and instead the Poissonian shot noise with $F \sim 1.0$ to 1.1 is observed. Moreover, mechanism to explain such $F$ suppression quantitatively has not been reported to date.

To calculate Fano factor, we examine full counting statistics including a spin-flip relaxation. (The details of the calculation are described in the Supplemental Material.) We note that previous theories have predicted the presence of super-Poissonian shot noise, which can explain previous experiments, but calculation including spin-flip relaxation has not been reported yet. This is understood because previous experiments intrinsically have smaller spin-flip relaxation, since they were performed for quantum dots fabricated from material with smaller spin-orbit interaction such as GaAs [15,16], CNT [18], and Si [17]. This is in contrast with our measurements of an InAs nanowire quantum dot with strong spin-orbit interaction, where the spin flip can play an important role.

The spin-flip relaxation in our model is described with a rate from the initial state $|n, m_i\rangle$ to final state $|n, m_f\rangle$ ($n$: total electron number, $m$: the spin configuration) given by

$$\frac{\gamma}{\hbar} |\langle n, m_f | (\widehat{\sigma_+} + \widehat{\sigma_-}) | n, m_i \rangle|^2 \theta \left( E_{n,m_i} - E_{n,m_f} \right),$$

where $\gamma$ is the strength of the spin-flip relaxation, $\hbar$ is the Planck constant, $\widehat{\sigma_\pm}$ is the spin ladder operator, $E_{n,m_{i(f)}}$ is the energy of $|n, m_{i(f)}\rangle$ states. (See the Supplemental Material). We consider that the spin flip occurs with the rate of the order of $\gamma/\hbar$. The schematic illustration that is helpful for intuitive understanding is shown in Fig. 4(d). Inelastic cotunneling is ordinarily represented by the process of (1)-(2)-(1). When there arises an electron bunching or dynamical channel blockade [13,14], in which multiple electron channels with different coupling strength with the leads interact with each other, the process of (2-1) and/or (2-2) are included. This consequently leads to effectively large $F$, i.e. super-Poissonian noise. On the other hand, the spin flip that we consider is described by (3), which acts to prevent dynamical channel blockade, leading to smaller $F$. In physical meaning, this spin-flip relaxation can be induced by spin-orbit interaction and phonon emission.

The simulation that were obtained with this effect is shown in Fig. 4(b), which is comparable to experimental results shown in Fig. 4(a). Here the density of states of electrodes are used as those of superconductor (normal metal) at zero (finite) fields. The $g$ factor to be used for Zeeman energy is ~ 3.3, which is determined to match with the size of the Coulomb diamonds. $g$ factor of bulk InAs is about 14, but it is reported to be smaller (2 ~ 10) in a quantum dot [37,38], depending on electron confinement potential. The left panels in Fig. 4(b) show simulation obtained at 0 and 0.9 T with $\gamma$ = 0 (without spin-flip relaxation effect). At zero field, $F$ ~ 4 at $eV_{sd} > \Delta E + 2\Delta_s$ for even parity, and for odd parity, $F$ is mostly 1 and partly reaching 1.3 in the corresponding area. In contrast, at 0.9 T where superconducting gap disappears and instead Zeeman gap occurs, $F$ ~ 3 (2) at even (odd) parity. These results indicate that the overall features are consistent with the qualitative schematic diagram shown in Fig. 4(c). On the other hand, the right panels in Fig. 4(b) show simulation obtained with $\gamma$ = 2 meV. Compared to $\gamma$ = 0 meV case, $F$ is rather small and is ~ 2 (~ 1) for even (odd) parity. This shows good match with our observation.

We next show $B$ dependence of experimental results and simulation in Fig. 4(e). For even occupation, $F$ is the largest at $B = 0$, approaching a constant value with increasing $B$. This trend does not depend on $\gamma$ value, but with increasing $\gamma$, $F$ becomes smaller at any magnetic field. In contrast, for odd occupation with $\gamma = 0$, $F$ becomes larger when $B$ is increased and Zeeman gap opens, finally reaching super-Poissonian shot noise. With increasing $\gamma$, $F$ drastically decreases and approaches Poissonian noise. By comparing our experiments with this simulation, we can quantitatively and coherently explain our results with $\gamma$ ~ 2 meV.

The value of $\gamma$ = 2 meV is about ten times larger than the spin-orbit interaction energy deduced for InAs nanowire quantum dot [37] and InAs self-assembled quantum dot [38], both of which were obtained at high magnetic field from excited-states spectroscopy [37] and Kondo effect [38]. On the other hand, $\gamma$ ~ 2 meV is of similar magnitude to the spin-orbit interaction energy given by $2\alpha_R E_F$ ($\alpha_R$: Rashba coupling constant, $E_F$: Fermi energy) obtained for gate-all-around metal-oxide InAs nanowire FET that we previously developed [30]. The reason for the large γ is unknown, but this is partly because our model employs a specific spin-flip relaxation process that impacts Fano factor, which is too simple to describe whole effect that can occur by spin-orbit interaction. There may also be other factors inducing spin-flip relaxation except the simple spin-orbit interaction. We also mention that, in our calculation, Fano factor in the Coulomb blockade regime depends on correlation between the spin-flip relaxation and dot-lead coupling, the details of which will be published elsewhere.

In our simulation to represent spin-flip relaxation that can be induced by the spin-orbit interaction, we use γ that is constant with $B$. Here it is interesting to compare our case with previous papers reporting that the spin-relaxation rate between Zeeman sub-levels is proportional to $B^5$ in the presence of the spin-orbit interaction [39]. Clear difference from our case is the totally different situation; previous report focuses on a GaAs quantum dot made in bulk two-dimensional electron gas and the $B^5$ relaxation rate results from the three-dimensional bulk phonons with wavelength longer than a quantum dot. Thus, we consider that such rapid relaxation with $B$ does not seem to be our case.

In conclusion, we carried out shot noise measurements for a hybrid device consisting of an InAs nanowire quantum dot and a superconductor. We observe the parity effect of the super-Poissonian and Poissonian alternation at zero field, and this effect disappears with increasing magnetic fields, demonstrating that shot noise approaches the Poissonian statistics. These experimental observations are coherently and quantitatively explained by spin-flip relaxation that can happen in the presence of strong spin-orbit interaction and phonon relaxation.

**Acknowledgement**

We thank H. Murofushi for technical support, and A. Iwasaki and Y. Tokura for helpful discussions.

This work was partially supported by JSPS KAKENHI Grants 20H02562, 17K05575, 18KK0385, and 20H01827.

* [keiko.takase.wa@hco.ntt.co.jp](keiko.takase.wa@hco.ntt.co.jp)

Figures

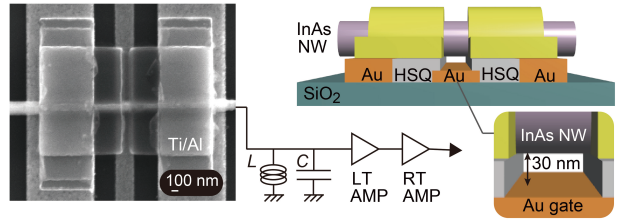

FIG. 1 (color online) A SEM image of our typical sample (top view) with an experimental setup for performing shot noise measurements. A schematic of a cross-section view of the sample is also shown.

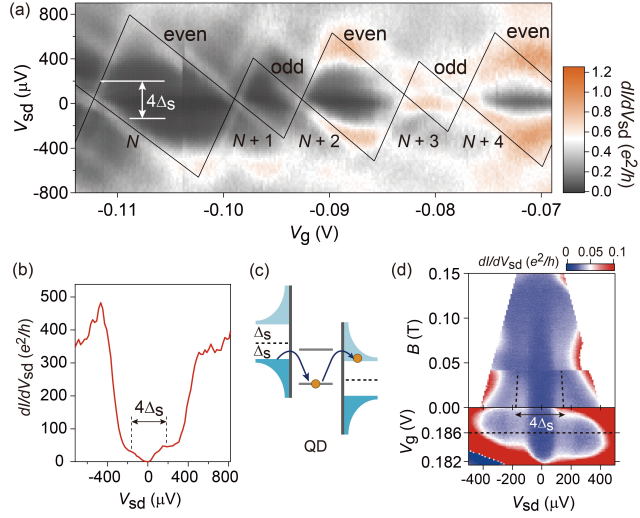

FIG. 2 (color online) (a) differential conductance $dI/dV_{sd}$ as a function of gate voltage $V_g$ and source drain voltage $V_{sd}$. Solid lines show the edges of Coulomb diamonds. (b) Cut lines of $dI/dV$ vs. $V_{sd}$ obtained at $V_g = -0.106$ V. The cotunneling peaks associated with superconducting electrodes are indicated. (c) Schematics of cotunneling. (d) Magnetic field dependence of the cotunneling peaks (top) and the Coulomb diamond at a zero field with the peaks inside (bottom). The data in (d) were obtained for a different Coulomb diamond in different measurement runs.

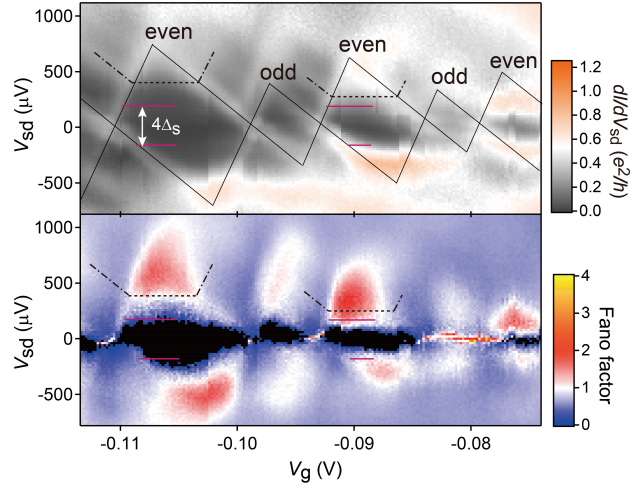

FIG. 3 (color online) $dI/dV_{sd}$ (top panel) and Fano factor (bottom panel) plotted as a function of $V_g$ and $V_{sd}$. Dashed-dotted lines outside the Coulomb diamonds are the guides indicating the sequential tunneling. The blacked out region in the bottom panel is the area where the electrical signals are below the sensitivity of our noise measurement system.

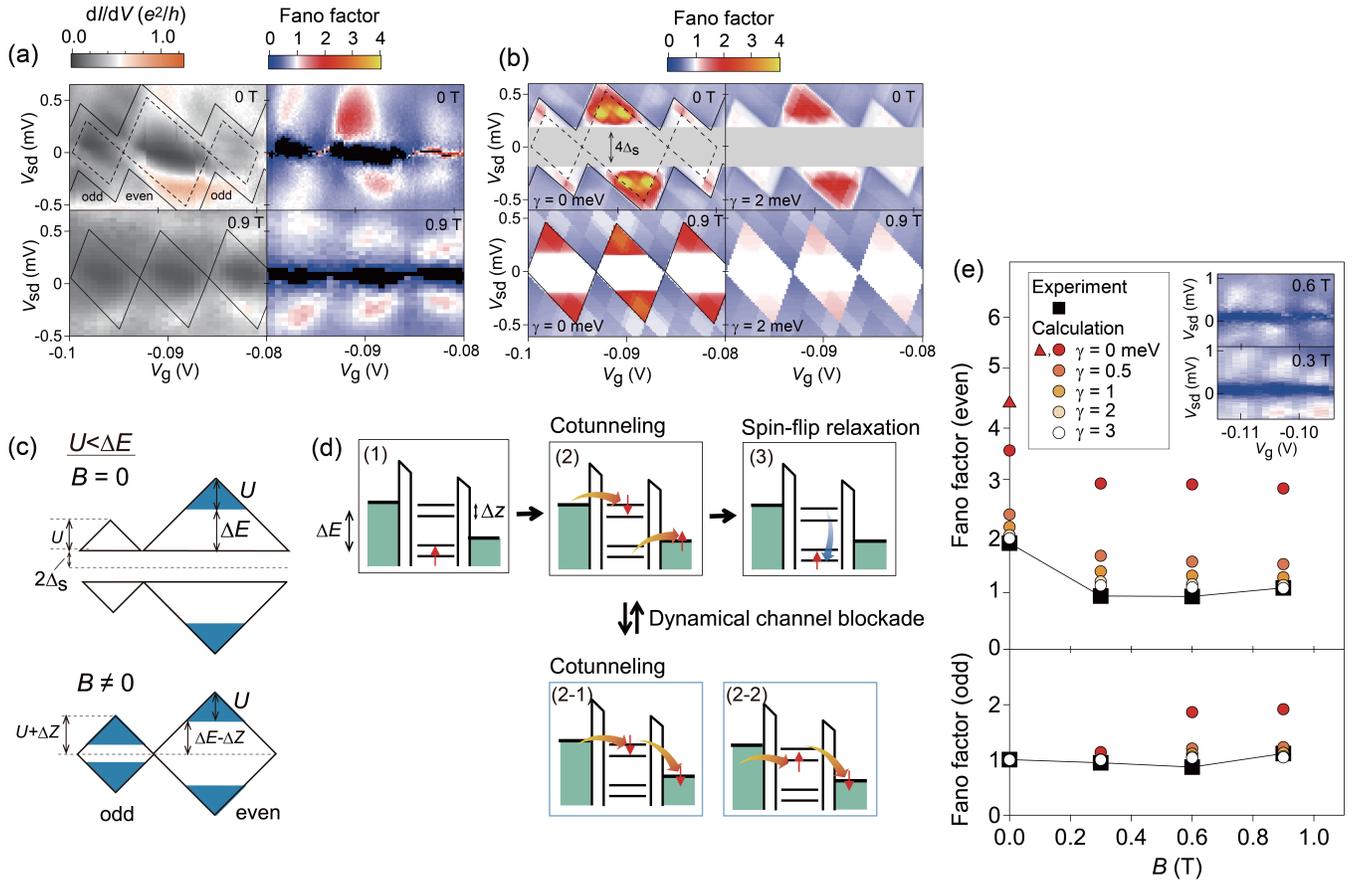

Fig. 4 (color online) (a) $dI/dV_{sd}$ (left panels) and Fano factor (right panels) plotted as a function of $V_g$ and $V_{sd}$, which are measured at 0 T and 0.9 T. (b) Simulation results using full-counting statistics. Panels for $\gamma = 2$ meV mostly reproduce the experimental results shown in Fig. 4(a). $V_{sd} < 2|\Delta_s|$ are filled in gray because measurements in a part of this area are below the sensitivity limit as is noted in Fig. 3. (c) Schematic diagram of the Coulomb diamonds at 0 T and finite magnetic fields. (d) Schematic illustration of the inelastic cotunneling involving dynamical channel blockade with illustration of the spin-flip relaxation process. (e) Fano factor as a function of magnetic field for the experiments and the simulations performed with different $\gamma$. For $\gamma = 0$ at zero field, circle (triangle) symbol represents the value at the center (the maximum value) in the area of the super-Poissonian shot noise. The inset shows the Fano factor mapping obtained at 0.3 T and 0.6 T.

# Supplemental Material for Parity-dependent shot noise and spin-flip relaxation process in a hybrid superconductor-nanowire quantum dot

Keiko Takase[1], Uasuhiro Utsumi[2], Yamato Ashikawa[1,3],
Guoqiang Zhang[1], Kouta Tateno[1], Yuma Okazaki[1] and Satoshi Sasaki[1]

[1]*NTT Basic Research Laboratories, NTT Corporation,*
*3-1 Morinosato-Wakamiya, Atsugi, Kanagawa 243-0198, Japan*
[2]*Department of Physics Engineering, Faculty of Engineering, Mie University, Tsu, Mie, 514-8507, Japan*
[3]*Department of Physics, Tohoku University, 6-3 Aramaki Aza Aoba, Aoba-ku, Sendai 980-8578, Japan*

**CONTENTS**

## I. INTRODUCTION

We show how to calculate the current noise of cotunneling processes in a quantum dot. To match with our experiments performed for the InAs quantum dot with superconducting electrodes, we consider a spinful quantum dot connected with superconducting leads. We examine the case from zero to finite magnetic fields, where the leads have changed from superconducting to normal states and furthermore the dot energy levels have been separated by the Zeeman energy. Moreover, being different from the previous theories, [2–6] our model includes a spin-flip relaxation process, which can manifest itself in the presence of strong spin-orbit interaction that is present in narrow-gap semiconductors.

We note that the previous theories predicting the presence of the super-Poissonian noise [2–6] are qualitatively consistent with previous experimental observations that are performed at zero field using CNT [7], GaAs [8, 9], and Si [10] quantum dots, while the values of the super-Poissonian noise are different from paper to paper due to the different assumptions, parameters, and models. We also note that the spin-orbit interaction in the previously used samples is much weaker than that of InAs, indicating that the spin-flip relaxation process that we consider in this paper was expected to be experimentally negligible.

To estimate the current noise, we calculate full-counting statistics in a modified master equation [13]. Our calculation technique itself is similar to Refs. 1 and 11, with the difference that we use the Rayleigh-Schrödinger perturbation theory to obtain higher-order current cumulants [16]. The tunneling rates used for the calculation are derived using Fermi golden rule. Even though this technique is less systematic than the real-time diagrammatic theory, [2–6] our method is simpler and is able to predict the occurrence of the super-Poissonian noise in the Coulomb blockade regime, as is consistent with the previous theoretical calculations. [1–6, 11] More importantly, we find that the spin-flip relaxation acts as disturbing the dynamical channel blockade that occurs in the cotunneling process, and find that this situation systematically suppresses the super-Poissonian noise. Consequently, our model coherently and quantitatively explain the values of the Fano factor experimentally obtained for our device from zero to finite magnetic fields (the comparisons between the experiments and the calculation are presented in the paper).

This supplementary information organizes as follows. Section II presents the model Hamiltonian. Section III shows the transition rates for the each process involved in the quantum dot tunneling. Section IV describes how to extract electrical current and current fluctuation. Section V shows simulation and also summarizes detailed parameters, which are used to reproduce our experiments in the paper.

## II. MODEL HAMILTONIAN

The Hamiltonian of our system is,

$$\hat{H} = \hat{H}_D + \sum_{r=L,R} \hat{H}_r + \hat{H}_T \,. \tag{1}$$

The dot Hamiltonian is,

$$\hat{H}_D = \sum_{\sigma=\uparrow,\downarrow} \sum_{s=\pm} \epsilon_s \hat{n}_{s\sigma} + \frac{U}{2} \sum_{s',\sigma' \neq s,\sigma} \hat{n}_{s'\sigma'} \hat{n}_{s\sigma} + \Delta Z \sum_{s=\pm} \frac{\hat{n}_{s\uparrow} - \hat{n}_{s\downarrow}}{2} \,. \tag{2}$$

The quantum dot contains two energy levels $\epsilon_+$ and $\epsilon_-$. $\hat{n}_{s\sigma} = \hat{d}^\dagger_{s\sigma} \hat{d}_{s\sigma}$ is the number operator of electron and $\hat{d}_{s\sigma}$ annihilate an electron with spin $\sigma$ in the level $s$. The lower energy level $\epsilon_-$ is tuned by the gate voltage and also affected by the source-drain bias voltage; $\epsilon_- = \beta(eV_g - eV_{g,0}) + \alpha eV_{sd}$, where $V_{g,0}$ is an offset of the gate voltage. We assume that the level spacing between the two levels is independent of the gate and source-drain bias voltages, $\epsilon_+ - \epsilon_- = \Delta E$. The Coulomb interaction energy is assumed to be constant $U$ and the Zeeman energy is given by $\Delta Z$.

The lead $r$ ($r = L, R$) is modeled by a non-interacting gas of quasi-particles.

$$\hat{H}_r = \sum_{k\sigma} \epsilon_{rk} \hat{a}^\dagger_{rk\sigma} \hat{a}_{rk\sigma} \,, \tag{3}$$

where $\hat{a}_{rk\sigma}$ annihilate a quasi-particle with spin $\sigma$ and wave number $k$ in the lead $r$. We write DOS for the superconducting electrodes as

$$\sum_k \delta(\epsilon - \epsilon_{rk}) = \rho_{0,r} \bar{\rho}_r(\epsilon) \,, \quad \bar{\rho}_r(\epsilon) = \mathrm{Re} \frac{1}{\sqrt{1 - \Delta_S^2/(\epsilon + i0^+)^2}} \,, \tag{4}$$

where $\Delta_S$ is the half of the superconducting gap and $0^+$ is a positive infinitesimal.

When the lead is varied from superconducting to metallic states with increasing magnetic fields, $\bar{\rho}_r(\epsilon) = 1$. The tunnel Hamiltonian is,

$$\hat{H}_T = \sum_{r=L,R} \sum_{\sigma=\uparrow,\downarrow} \sum_{s=\pm} \sum_k J_{rs} \hat{d}^\dagger_{s\sigma} \hat{a}_{rk\sigma} + \text{H.c.}. \tag{5}$$

The dot Hamiltonian $\hat{H}_D$ is diagonalized and there are 16 dots states (see Fig. S1). When the quantum dot contains $n$ electrons, there are $M(n) = \binom{4}{n}$ eigen states, which we denote $|n, m\rangle$ ($m = 1, \cdots, M(n)$). The eigen energies are $E_{n,m}$. In total, there are $\sum_{n=0}^{4} M(n) = N = 2^4$ eigen states;

$$|0, 1\rangle = |0\rangle, \tag{6}$$

$$|1, 1\rangle = d^\dagger_{-\uparrow}|0\rangle, \quad |1, 2\rangle = d^\dagger_{-\downarrow}|0\rangle, \quad |1, 3\rangle = d^\dagger_{+\uparrow}|0\rangle, \quad |1, 4\rangle = d^\dagger_{+\downarrow}|0\rangle, \tag{7}$$

$$|2, 1\rangle = d^\dagger_{-\downarrow} d^\dagger_{-\uparrow}|0\rangle, \quad |2, 2\rangle = d^\dagger_{+\uparrow} d^\dagger_{-\uparrow}|0\rangle, \quad |2, 3\rangle = d^\dagger_{+\downarrow} d^\dagger_{-\uparrow}|0\rangle, \quad |2, 4\rangle = d^\dagger_{+\uparrow} d^\dagger_{-\downarrow}|0\rangle, \tag{8}$$

$$|2, 5\rangle = d^\dagger_{+\downarrow} d^\dagger_{-\downarrow}|0\rangle, \quad |2, 6\rangle = d^\dagger_{+\downarrow} d^\dagger_{+\uparrow}|0\rangle, \tag{9}$$

$$|3, 1\rangle = d^\dagger_{+\uparrow} d^\dagger_{-\downarrow} d^\dagger_{-\uparrow}|0\rangle, \quad |3, 2\rangle = d^\dagger_{+\downarrow} d^\dagger_{-\downarrow} d^\dagger_{-\uparrow}|0\rangle, \quad |3, 3\rangle = d^\dagger_{+\downarrow} d^\dagger_{+\uparrow} d^\dagger_{-\uparrow}|0\rangle, \quad |3, 4\rangle = d^\dagger_{+\downarrow} d^\dagger_{+\uparrow} d^\dagger_{-\downarrow}|0\rangle, \tag{10}$$

$$|4, 1\rangle = d^\dagger_{+\downarrow} d^\dagger_{+\uparrow} d^\dagger_{-\downarrow} d^\dagger_{-\uparrow}|0\rangle. \tag{11}$$

The eigen energies are,

$$E_{0,1} = 0, \tag{12}$$

$$\{E_{1,m}\} = \{\epsilon_- + \Delta Z/2, \epsilon_- - \Delta Z/2, \epsilon_+ + \Delta Z/2, \epsilon_+ + \Delta Z/2\}, \tag{13}$$

$$\{E_{2,m}\} = \{2\epsilon_- + U, \epsilon_+ + \epsilon_- + \Delta Z + U, \epsilon_+ + \epsilon_- + U, \epsilon_+ + \epsilon_- + U, \epsilon_+ + \epsilon_- - \Delta Z + U, 2\epsilon_+ + U\}, \tag{14}$$

$$\{E_{3,m}\} = \{2\epsilon_- + \epsilon_+ + 3U + \Delta Z/2, 2\epsilon_- + \epsilon_+ + 3U - \Delta Z/2, \epsilon_- + 2\epsilon_+ + 3U + \Delta Z/2, \epsilon_- + 2\epsilon_+ + 3U - \Delta Z/2\}, \tag{15}$$

$$E_{4,1} = 2(\epsilon_- + \epsilon_+) + 6U. \tag{16}$$

Then by using the projection operator, the Hamiltonians are rewritten as,

$$\hat{H}_D = \sum_n \sum_{m=1}^{M(n)} E_{n,m} |n, m\rangle\langle n, m|, \tag{17}$$

$$\hat{H}_T = \sum_{rk\sigma} \sum_n \sum_{m'=1}^{M(n+1)} \sum_{m=1}^{M(n)} \sqrt{\frac{1}{2\pi \rho_{r,0}}} \left(\boldsymbol{g}^{(n+1,n)}_{r,\sigma}\right)_{m'm} \hat{a}_{rk\sigma} |n+1, m'\rangle\langle n, m| + \text{H.c.}, \tag{18}$$

$$\left(\boldsymbol{g}^{(n+1,n)}_{r,\sigma}\right)_{m'm} = \sum_{s''=\pm} \sqrt{G_{rs''}} \langle n+1, m'|\hat{d}^\dagger_{s''\sigma}|n, m\rangle. \tag{19}$$

Here we introduce the tunnel coupling strength as, $G_{r\pm} = 2\pi\rho_{0,r} J^2_{r\pm}$. Here $\boldsymbol{g}^{(n+1,n)}_{r,\sigma}$ is a $M(n+1) \times M(n)$ matrices. The explicit forms are,

$$\boldsymbol{g}^{(1,0)}_{r,\uparrow} = \begin{pmatrix} \sqrt{G_{r-}} \\ 0 \\ \sqrt{G_{r+}} \\ 0 \end{pmatrix}, \quad \boldsymbol{g}^{(1,0)}_{r,\downarrow} = \begin{pmatrix} 0 \\ \sqrt{G_{r-}} \\ 0 \\ \sqrt{G_{r+}} \end{pmatrix}. \tag{20}$$

$$\boldsymbol{g}^{(2,1)}_{r,\uparrow} = \begin{pmatrix} 0 & -\sqrt{G_{r-}} & 0 & 0 \\ \sqrt{G_{r+}} & 0 & -\sqrt{G_{r-}} & 0 \\ 0 & 0 & 0 & -\sqrt{G_{r-}} \\ 0 & \sqrt{G_{r+}} & 0 & 0 \\ 0 & 0 & 0 & 0 \\ 0 & 0 & 0 & -\sqrt{G_{r+}} \end{pmatrix}, \quad \boldsymbol{g}^{(2,1)}_{r,\downarrow} = \begin{pmatrix} \sqrt{G_{r-}} & 0 & 0 & 0 \\ 0 & 0 & 0 & 0 \\ \sqrt{G_{r+}} & 0 & 0 & 0 \\ 0 & 0 & -\sqrt{G_{r-}} & 0 \\ 0 & \sqrt{G_{r+}} & 0 & -\sqrt{G_{r-}} \\ 0 & 0 & \sqrt{G_{r+}} & 0 \end{pmatrix}, \tag{21}$$

$$\left(\boldsymbol{g}_{r,\uparrow/\downarrow}^{(3,2)}\right)_{m'm} = \mp \left(\boldsymbol{g}_{r,\uparrow/\downarrow}^{(2,1)}\right)_{5-m\,7-m}, \quad \left(\boldsymbol{g}_{r,\uparrow/\downarrow}^{(4,3)}\right)_{m} = \mp \left(\boldsymbol{g}_{r,\uparrow/\downarrow}^{(1,0)}\right)_{5-m}. \tag{22}$$

The matrix elements in above matrices can be automatically generated by exploiting the Jordan-Wigner transform [12],

$$d^\dagger_{-\uparrow} = \sigma_z \otimes \sigma_z \otimes \sigma_z \otimes \sigma_+, \quad d^\dagger_{-\downarrow} = \sigma_z \otimes \sigma_z \otimes \sigma_+ \otimes \sigma_0, \quad d^\dagger_{+\uparrow} = \sigma_z \otimes \sigma_+ \otimes \sigma_0 \otimes \sigma_0, \quad d^\dagger_{-\uparrow} = \sigma_+ \otimes \sigma_0 \otimes \sigma_0 \otimes \sigma_0, \tag{23}$$

where

$$\sigma_z = \begin{pmatrix} 1 & 0 \\ 0 & -1 \end{pmatrix}, \quad \sigma_+ = \begin{pmatrix} 0 & 1 \\ 0 & 0 \end{pmatrix}, \quad \sigma_0 = \begin{pmatrix} 1 & 0 \\ 0 & 1 \end{pmatrix}, \tag{24}$$

are Pauli matrices and the unit matrix.

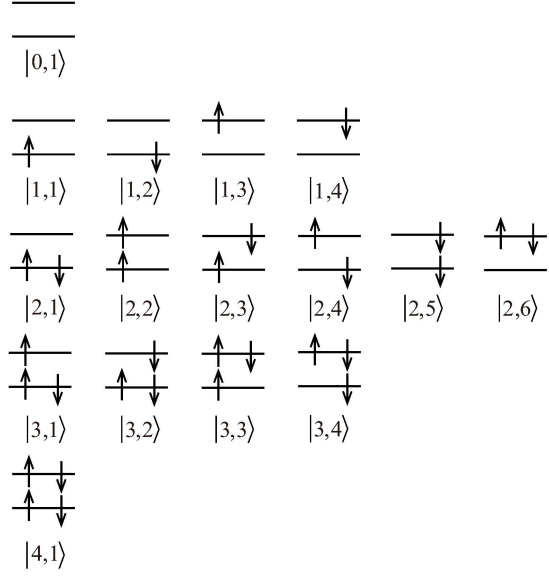

FIG. S1. 16 dot states considered in our calculation

### III.  TRANSITION RATES

The transition rates are calculated within the Fermi golden rule.[13] In the following, we consider the zero temperature limit, since the measurements temperature is low enough ( $\sim$50 mK) to ignore thermal effect.

#### A.  Sequential tunneling process

The transition rate from $|n, m_i\rangle$ to $\hat{a}_{rk\sigma}|n+1, m_f\rangle$ is given by,

$$\Gamma^{\text{seq}}_{r;n+1,m_f \leftarrow n,m_i} = \sum_{k\sigma} \frac{2\pi}{\hbar} \left|\langle n+1, m_f|\hat{a}^\dagger_{rk\sigma}\hat{H}_T|n, m_i\rangle\right|^2 \delta(E_{n+1,m_f} - \epsilon_{rk} - E_{n,m_i})\,\theta(\mu_r - \epsilon_{rk}) \tag{25}$$

$$= \frac{1}{\hbar} \sum_\sigma \left(\boldsymbol{g}_{r,\sigma}^{(n+1,n)}\right)^2_{m_f,m_i} \bar{\rho}_r(E_{n+1,m_f} - E_{n,m_i} - \mu_r)\,\theta(\mu_r - E_{n+1,m_f} + E_{n,m_i}), \tag{26}$$

where the Heaviside step function is $\theta(x) = 1$ for $x > 0$ and $\theta(x) = 0$ otherwise. Similarly, the transition rate to the opposite direction is calculated as,

$$\Gamma^{\text{seq}}_{r;n,m_i \leftarrow n+1,m_f} = \frac{1}{\hbar} \sum_\sigma \left(\boldsymbol{g}_{r,\sigma}^{(n+1,n)}\right)^2_{m_f,m_i} \bar{\rho}_r(E_{n+1,m_f} - E_{n,m_i} - \mu_r)\,\theta(E_{n+1,m_f} - E_{n,m_i} - \mu_r). \tag{27}$$

### B. Cotunneling process

Next we consider the second order process. For simplicity, we only consider the cotunneling processes, in which the quantum dot charge state does not change after the particle tunneling. In other words, we do not consider two particle emission (absorption) process from (to) the quantum dot. The transition rate of the cotunneling process from $|n, m_i\rangle$ to $\hat{a}^\dagger_{rk\sigma} \hat{a}_{r'k'\sigma'} |n, m_f\rangle$ is given by,

$$\Gamma^{\text{cot}}_{r,n,m_f \leftarrow r',n,m_i} = \sum_{k\sigma} \sum_{k'\sigma'} \frac{2\pi}{\hbar} \left| \langle n, m_f | \hat{a}^\dagger_{r'k'\sigma'} \hat{a}_{rk\sigma} \hat{H}_T \frac{1}{E_{n,m_i} - \hat{H}_0 - i\hbar\hat{\Gamma}/2 - i\bar{\eta}} \hat{H}_T | n, m_i \rangle \right|^2$$
$$\times \delta(E_{n,m_f} + \epsilon_{rk} - \epsilon_{r'k'} - E_{n,m_i}) \theta(\mu_r - \epsilon_{rk\sigma}) \theta(\epsilon_{r'k'\sigma'} - \mu_{r'}), \tag{28}$$

where we introduced the decay rate of the intermediate state $|n, m\rangle$ [14],

$$\hat{\Gamma} = \sum_n \sum_{m=1}^{M(n)} (\Gamma_{n,m} + \eta) |n,m\rangle \langle n,m|, \quad \Gamma_{n,m} = \sum_r \sum_\pm \sum_{m'=1}^{M(n\pm 1)} \Gamma^{\text{seq}}_{r;n\pm 1,m' \leftarrow n,m}. \tag{29}$$

Here $\eta$ is a small broadening introduced by hand, which induces a decay in the intermediate states in cotunneling process. Equation (28) is calculated as

$$\Gamma^{\text{cot}}_{r',n,m_f \leftarrow r,n,m_i} = \frac{1}{2\pi\hbar} \sum_{\sigma\sigma'} \int d\epsilon \theta(\mu_{r'} + E_{n,m_i} - E_{n,m_f} - \epsilon) \theta(\epsilon - \mu_r) \left| \sum_{m=1}^{M(n+1)} \frac{\left(g^{(n+1,n)}_{r,\sigma}\right)_{m,m_f} \left(g^{(n+1,n)}_{r',\sigma'}\right)_{m,m_i}}{E_{n,m_f} - E_{n+1,m} + \epsilon - i\hbar\Gamma_{n+1,m}/2 - i\hbar\eta} \right.$$
$$\left. - \sum_{m=1}^{M(n-1)} \frac{\left(g^{(n,n-1)}_{r',\sigma'}\right)_{m_f,m} \left(g^{(n,n-1)}_{r,\sigma}\right)_{m_i,m}}{E_{n,m_i} - E_{n-1,m} - \epsilon - i\hbar\Gamma_{n-1,m}/2 - i\hbar\eta} \right|^2 \tag{30}$$

$$= \frac{\hbar}{2\pi} \sum_{\sigma,\sigma'} \sum_{j,j'=1}^{N(n+1)+N(n-1)} \frac{b^{\sigma\sigma'}_j b^{\sigma\sigma'}_{j'}}{a_j - a^*_{j'}} [F(\bar{\mu}_{r'}, \bar{\mu}_r, a_j) - F(\bar{\mu}_{r'}, \bar{\mu}_r, a^*_{j'})], \tag{31}$$

where

$$F(\mu', \mu, \alpha) = \theta(\mu' - \mu) \int_\mu^{\mu'} dz \frac{\bar{\rho}(z-\mu)\bar{\rho}(z-\mu')}{z - \alpha}. \tag{32}$$

Here $a_j$ and $b^{\sigma,\sigma'}_j$ are components of $M(n-1) + M(n+1)$ dimensional vectors;

$$\boldsymbol{a} = (\bar{E}_{n+1,1}, \cdots, \bar{E}_{n+1,M(n+1)}, -\bar{E}_{n-1,1}, \cdots, -\bar{E}_{n-1,M(n-1)}), \tag{33}$$

$$\boldsymbol{b}^{\sigma,\sigma'} = \left( \left(g^{(n+1,n)}_{r,\sigma}\right)_{1,m_f} \left(g^{(n+1,n)}_{r',\sigma'}\right)_{1,m_i}, \cdots, \left(g^{(n+1,n)}_{r,\sigma}\right)_{M(n+1),m_f} \left(g^{(n+1,n)}_{r',\sigma'}\right)_{M(n+1),m_i}, \right.$$
$$\left. \left(g^{(n,n-1)}_{r',\sigma'}\right)_{m_f,1} \left(g^{(n,n-1)}_{r,\sigma}\right)_{m_i,1}, \cdots, \left(g^{(n,n-1)}_{r',\sigma'}\right)_{m_f,M(n-1)} \left(g^{(n,n-1)}_{r,\sigma}\right)_{m_i,M(n-1)} \right). \tag{34}$$

where $\bar{\mu}_{r'} = \mu_{r'} - (E_{n,m_f} - E_{n,m_i})/2$ and $\bar{\mu}_r = \mu_r + (E_{n,m_f} - E_{n,m_i})/2$. The complex energy is $\bar{E}_{n',m} = E_{n',m} - (E_{n,m_f} + E_{n,m_i})/2 + i\hbar(\Gamma_{n',m}/2 + \eta)$. The off diagonal component $f^{\sigma\sigma'}_{j,j'}$ ($j \neq j'$) represents the interference of two cotunneling processes.

### C. Spin-relaxation rate

In addition, we introduce a phenomenological spin relaxation. We assume the following spin-relaxation rate,

$$\Gamma^{\text{relax}}_{n,m_f \leftarrow n,m_i} = \frac{\gamma}{\hbar} |\langle n, m_f | (\hat{\sigma}_+ + \hat{\sigma}_-) | n, m_i \rangle|^2 \theta(E_{n,m_i} - E_{n,m_f}), \quad \hat{\sigma}_+ = \sum_{s,s'=\pm} \hat{d}^\dagger_{s\uparrow} \hat{d}_{s'\downarrow}, \quad \hat{\sigma}_- = \hat{\sigma}^\dagger_+, \tag{35}$$

where $\gamma$ is the strength of spin-flip relaxation. Such transition process may be induced by the spin-orbit interaction and the electron-phonon coupling.

## IV. AVERAGE CURRENT AND CURRENT NOISE

### A. Full-counting statistics

By exploiting the transition rates, we can calculate the full-counting statistics in the master equation approach [15]. Let $p_{n,m}$ be the probability to find the state $|n,m\rangle$. Then we introduce a $N$ dimensional vector $\boldsymbol{p}^T = (\boldsymbol{p}_0^T, \boldsymbol{p}_1^T, \boldsymbol{p}_2^T, \boldsymbol{p}_3^T, \boldsymbol{p}_4^T)$, where $\boldsymbol{p}_n = (p_{n,1}, \cdots, p_{n,M(n)})^T$. The system can be described with the modified master equation,

$$\frac{d}{dt}\boldsymbol{p}(t) = \boldsymbol{L}(\chi)\boldsymbol{p}(t),\tag{36}$$

where $\boldsymbol{L}(\chi)$ is the transition rate matrix modified with the counting filed $\chi$. The probability distribution of transmitted charge $P_\tau(q)$ and its characteristic function $\mathcal{Z}(\chi)$ are,

$$\sum_q P_\tau(q)e^{i\chi q} = \mathcal{Z}(\chi) = \boldsymbol{e}^T e^{\boldsymbol{L}(\chi)\tau}\boldsymbol{p}(0),\tag{37}$$

where $\boldsymbol{e}^T = (1, \cdots, 1)$. The transition rate matrix is

$$\begin{aligned}\boldsymbol{L}(\chi) =& \boldsymbol{L}_{L,+}^{\text{seq}}e^{i\chi/2} + \boldsymbol{L}_{L,-}^{\text{seq}}e^{-i\chi/2} + \boldsymbol{L}_{R,+}^{\text{seq}}e^{-i\chi/2} + \boldsymbol{L}_{R,-}^{\text{seq}}e^{i\chi/2} \\ &+ \boldsymbol{L}_{R\leftarrow L}^{\text{cot}}e^{i\chi} + \boldsymbol{L}_{L\leftarrow L}^{\text{cot}} + \boldsymbol{L}_{R\leftarrow R}^{\text{cot}} + \boldsymbol{L}_{L\leftarrow R}^{\text{cot}}e^{-i\chi} + \boldsymbol{L}^{\text{relax}} - \boldsymbol{L}_{\text{decay}}\,.\end{aligned}\tag{38}$$

The transition matrices associated with the sequential tunneling processes are

$$\boldsymbol{L}_{r,+}^{\text{seq}} = \begin{bmatrix} 0 & 0 & 0 & 0 & 0 \\ \boldsymbol{L}_{r;1\leftarrow 0}^{\text{seq}} & 0 & 0 & 0 & 0 \\ 0 & \boldsymbol{L}_{r;2\leftarrow 1}^{\text{seq}} & 0 & 0 & 0 \\ 0 & 0 & \boldsymbol{L}_{r;3\leftarrow 2}^{\text{seq}} & 0 & 0 \\ 0 & 0 & 0 & \boldsymbol{L}_{r;4\leftarrow 3}^{\text{seq}} & 0 \end{bmatrix}, \quad \boldsymbol{L}_{r,-}^{\text{seq}} = \begin{bmatrix} 0 & \boldsymbol{L}_{r;0\leftarrow 1}^{\text{seq}} & 0 & 0 & 0 \\ 0 & 0 & \boldsymbol{L}_{r;1\leftarrow 2}^{\text{seq}} & 0 & 0 \\ 0 & 0 & 0 & \boldsymbol{L}_{r;2\leftarrow 3}^{\text{seq}} & 0 \\ 0 & 0 & 0 & 0 & \boldsymbol{L}_{r;3\leftarrow 4}^{\text{seq}} \\ 0 & 0 & 0 & 0 & 0 \end{bmatrix}.\tag{39}$$

The component of a sub-matrix is, $(\boldsymbol{L}_{r;n\pm 1\leftarrow n}^{\text{seq}})_{m,m'} = \Gamma_{r;n\pm 1,m\leftarrow n,m'}^{\text{seq}}$. The transition matrices for cotunneling processes and spin-relaxation processes are block diagonal as,

$$\boldsymbol{L}_{r'\leftarrow r}^{\text{cot}} = \begin{bmatrix} \boldsymbol{L}_{0,r'\leftarrow 0,r}^{\text{cot}} & 0 & 0 & 0 & 0 \\ 0 & \boldsymbol{L}_{1,r'\leftarrow 1,r}^{\text{cot}} & 0 & 0 & 0 \\ 0 & 0 & \boldsymbol{L}_{2,r'\leftarrow 2,r}^{\text{cot}} & 0 & 0 \\ 0 & 0 & 0 & \boldsymbol{L}_{3,r'\leftarrow 3,r}^{\text{cot}} & 0 \\ 0 & 0 & 0 & 0 & \boldsymbol{L}_{4,r'\leftarrow 4,r}^{\text{cot}} \end{bmatrix}, \quad \boldsymbol{L}^{\text{relax}} = \begin{bmatrix} 0 & 0 & 0 & 0 & 0 \\ 0 & \boldsymbol{L}_1^{\text{relax}} & 0 & 0 & 0 \\ 0 & 0 & \boldsymbol{L}_2^{\text{relax}} & 0 & 0 \\ 0 & 0 & 0 & \boldsymbol{L}_3^{\text{relax}} & 0 \\ 0 & 0 & 0 & 0 & 0 \end{bmatrix}.\tag{40}$$

The components of a sub-matrix are $(\boldsymbol{L}_{n,r'\leftarrow n,r}^{\text{cot}})_{m,m'} = \Gamma_{r',n,m\leftarrow r,n,m'}^{\text{cot}}$ and $(\boldsymbol{L}_n^{\text{relax}})_{m,m'} = \Gamma_{n,m\leftarrow n,m'}^{\text{relax}}$. The diagonal matrix $\boldsymbol{L}_{\text{decay}}$ satisfies, $(\boldsymbol{L}_{\text{decay}})_{j,j'} = \delta_{j,j'}\sum_{\ell\neq j}\boldsymbol{L}(0)_{\ell,j}$ to ensure the conservation of the probability as $\boldsymbol{e}^T\boldsymbol{L}(0) = 0$.

### B. Rayleigh-Schrödinger perturbation theory

In the limit of the long measurement time $\tau \to \infty$, the characteristic function is dominated by the eigenvalue of $\boldsymbol{L}(\chi)$ with the maximum real part $\Lambda(\chi)$ as $\mathcal{Z}(\chi) \approx e^{\Lambda(\chi)\tau}$. The $n$-th current cumulant in the steady state is,

$$\mathcal{C}_n = \partial_{i\chi}^n \Lambda(\chi)\big|_{i\chi=0}\,.\tag{41}$$

We extend the Rayleigh-Schrödinger perturbation theory [13] to the non-Hermitian matrix $\boldsymbol{L}$ case to obtain the second cumulant from a modified master equation of full counting statistics [16]. This method is simple and suitable for the numerical calculation of higher order current cumulants. The transition rate matrix can be expanded in powers of $i\chi$;

$$\boldsymbol{L} = \boldsymbol{L}^{(0)} + i\chi\boldsymbol{L}^{(1)} + (i\chi)^2\boldsymbol{L}^{(2)} + \cdots\,.\tag{42}$$

We solve the eigenvalue equation,

$$\boldsymbol{L}\boldsymbol{u}_j = \Lambda_j \boldsymbol{u}_j\,, \tag{43}$$

up to $i\chi^2$ accuracy;

$$\Lambda_j = \Lambda_j^{(0)} + i\chi \Lambda_j^{(1)} + (i\chi)^2 \Lambda_j^{(2)} + \cdots. \tag{44}$$

By comparing it with Eq. (41), we observe the first and second cumulants are related to the first and second order corrections of the eigenvalue as $\mathcal{C}_1 = \Lambda_j^{(1)}$ and $\mathcal{C}_2 = 2\Lambda_j^{(2)}$. If we regard $\boldsymbol{L}$ as the 'Hamiltonian' and $\Lambda_j$ as the 'eigenvalue', our problem is to perform the second order Rayleigh-Schrodinger perturbation theory [13]. Since $\boldsymbol{L}$ is non-Hermitian, we also have to consider the left eigenvector $\boldsymbol{v}_j^T$ satisfying,

$$\boldsymbol{L}^T \boldsymbol{v}_j = \Lambda_j \boldsymbol{v}_j\,. \tag{45}$$

The left and right eigenvectors share the same eigenvalues since the characteristic equations are the same; $\det(\Lambda \mathbf{1} - \boldsymbol{L}) = \det(\Lambda \mathbf{1} - \boldsymbol{L}^T) = 0$. By exploiting Eqs. (43) and (45), we derive $(\Lambda_j - \Lambda_i)\boldsymbol{v}_i^T \boldsymbol{u}_j = 0$. Therefore, if there are no degenerate eigenvalues, left and right eigenvectors satisfy,

$$\boldsymbol{v}_i^T \boldsymbol{u}_j = 0\,, \quad (i \neq j)\,. \tag{46}$$

We expand the eigenvectors in powers of $i\chi$ as,

$$\boldsymbol{u}_j = \boldsymbol{u}_j^{(0)} + i\chi \boldsymbol{u}_j^{(1)} + (i\chi)^2 \boldsymbol{u}_j^{(2)} + \cdots\,, \quad \boldsymbol{v}_i^{(0)T} \boldsymbol{u}_j^{(n)} = 0\,, \quad (n = 1, 2, \cdots)\,, \quad i \neq j\,. \tag{47}$$

Then by substituting Eqs. (42), (44) and (47) into Eq. (43), we obtain

$$(\boldsymbol{L}^{(0)} - \Lambda_j^{(0)})\boldsymbol{u}_j^{(0)} = 0\,, \tag{48}$$

$$(\boldsymbol{L}^{(0)} - \Lambda_j^{(0)})\boldsymbol{u}_j^{(1)} + (\boldsymbol{L}^{(1)} - \Lambda_j^{(1)})\boldsymbol{u}_j^{(0)} = 0\,, \tag{49}$$

$$(\boldsymbol{L}^{(0)} - \Lambda_j^{(0)})\boldsymbol{u}_j^{(2)} + (\boldsymbol{L}^{(1)} - \Lambda_j^{(1)})\boldsymbol{u}_j^{(1)} + (\boldsymbol{L}^{(2)} - \Lambda_j^{(2)})\boldsymbol{u}_j^{(0)} = 0\,. \tag{50}$$

By taking the inner product of $\boldsymbol{v}_j^{(0)T}$ with Eq. (49) and by paying attention to the orthogonality condition (46), we obtain

$$\Lambda_j^{(1)} = \frac{\boldsymbol{v}_j^{(0)T} \boldsymbol{L}^{(1)} \boldsymbol{u}_j^{(0)}}{\boldsymbol{v}_j^{(0)T} \boldsymbol{u}_j^{(0)}}\,. \tag{51}$$

By taking the inner product of $\boldsymbol{v}_k^{(0)T}$ ($k \neq j$) with Eq. (49), we obtain $\boldsymbol{v}_k^{(0)T} \boldsymbol{u}_j^{(1)} = -\boldsymbol{v}_k^{(0)T} \boldsymbol{L}^{(1)} \boldsymbol{u}_j^{(0)}/(\Lambda_k^{(0)} - \Lambda_j^{(0)})$. We rewrite this expression by using a regular matrix $\boldsymbol{V}$ and the pseudo inverse matrix $\boldsymbol{Q}_j$,

$$\boldsymbol{V} = (\boldsymbol{v}_1^{(0)}, \cdots, \boldsymbol{v}_N^{(0)})\,, \quad \boldsymbol{Q}_j = \mathrm{diag}\left(\frac{1}{\Lambda_1^{(0)} - \Lambda_j^{(0)}}, \cdots, \frac{1}{\Lambda_{j-1}^{(0)} - \Lambda_j^{(0)}}, 0, \frac{1}{\Lambda_{j+1}^{(0)} - \Lambda_j^{(0)}}, \cdots, \frac{1}{\Lambda_n^{(0)} - \Lambda_j^{(0)}}\right)\,, \tag{52}$$

as,

$$\boldsymbol{V}^T \boldsymbol{u}_j^{(1)} = -\boldsymbol{Q}_j \boldsymbol{V}^T \boldsymbol{L}^{(1)} \boldsymbol{u}_j^{(0)}\,. \tag{53}$$

Then by taking the inner product of $\boldsymbol{v}_j^{(0)T}$ with Eq. (50) and by paying attention to the conditions Eq. (46) and Eq. (53), we obtain,

$$\Lambda_j^{(2)} = \frac{\boldsymbol{v}_j^{(0)T} \left[\boldsymbol{L}^{(2)} \boldsymbol{u}_j^{(0)} - (\boldsymbol{L}^{(1)} - \Lambda_j^{(1)})\boldsymbol{u}_j^{(1)}\right]}{\boldsymbol{v}_j^{(0)T} \boldsymbol{u}_j^{(0)}} = \frac{\boldsymbol{v}_j^{(0)T} \left[\boldsymbol{L}^{(2)} - (\boldsymbol{L}^{(1)} - \Lambda_j^{(1)})\left(\boldsymbol{V}^T\right)^{-1} \boldsymbol{Q}_j \boldsymbol{V}^T \boldsymbol{L}^{(1)}\right] \boldsymbol{u}_j^{(0)}}{\boldsymbol{v}_j^{(0)T} \boldsymbol{u}_j^{(0)}}\,, \tag{54}$$

We set the zero eigenvalue of $\boldsymbol{L}^{(0)} = \boldsymbol{L}(0)$ as the first component $\Lambda_1^{(0)} = 0$. The corresponding eigenvector is the steady state distribution probability $\boldsymbol{p}_{\mathrm{st}} \propto \boldsymbol{u}_1$. Since the steady state is expected to be unique, $\boldsymbol{v}_k^T \boldsymbol{u}_1 = 0$ for $k \neq 1$, from Eqs. (51) and (54), we obtain the first and second cumulants as,

$$\mathcal{C}_1 = \Lambda_1^{(1)} = \frac{\boldsymbol{v}_1^{(0)T} \boldsymbol{L}^{(1)} \boldsymbol{u}_1^{(0)}}{\boldsymbol{v}_1^{(0)T} \boldsymbol{u}_1^{(0)}}\,, \quad \mathcal{C}_2 = 2\Lambda_1^{(2)} = 2\frac{\boldsymbol{v}_1^{(0)T} \left[\boldsymbol{L}^{(2)} - \boldsymbol{L}^{(1)} \left(\boldsymbol{V}^T\right)^{-1} \boldsymbol{Q}_1 \boldsymbol{V}^T \boldsymbol{L}^{(1)}\right] \boldsymbol{u}_1^{(0)}}{\boldsymbol{v}_1^{(0)T} \boldsymbol{u}_1^{(0)}}\,, \tag{55}$$

where $\boldsymbol{Q}_1 = \mathrm{diag}\left(0, 1/\Lambda_2^{(0)}, \cdots, 1/\Lambda_N^{(0)}\right)$.

## V. SIMULATION AND PARAMETERS USED TO REPRODUCE EXPERIMENTS

### A. Simulation obtained with various parameters

Based on Eq. (55), the Fano factors ($F$) given by $\mathcal{C}_2/\mathcal{C}_1$ are calculated numerically with various parameters. Figure S2 shows $F$ as a function of source-drain voltage ($V_{sd}$) and gate voltage ($V_g$). (Here we focus on the case at zero field. The results in the presence of the magnetic fields are described in the paper.)

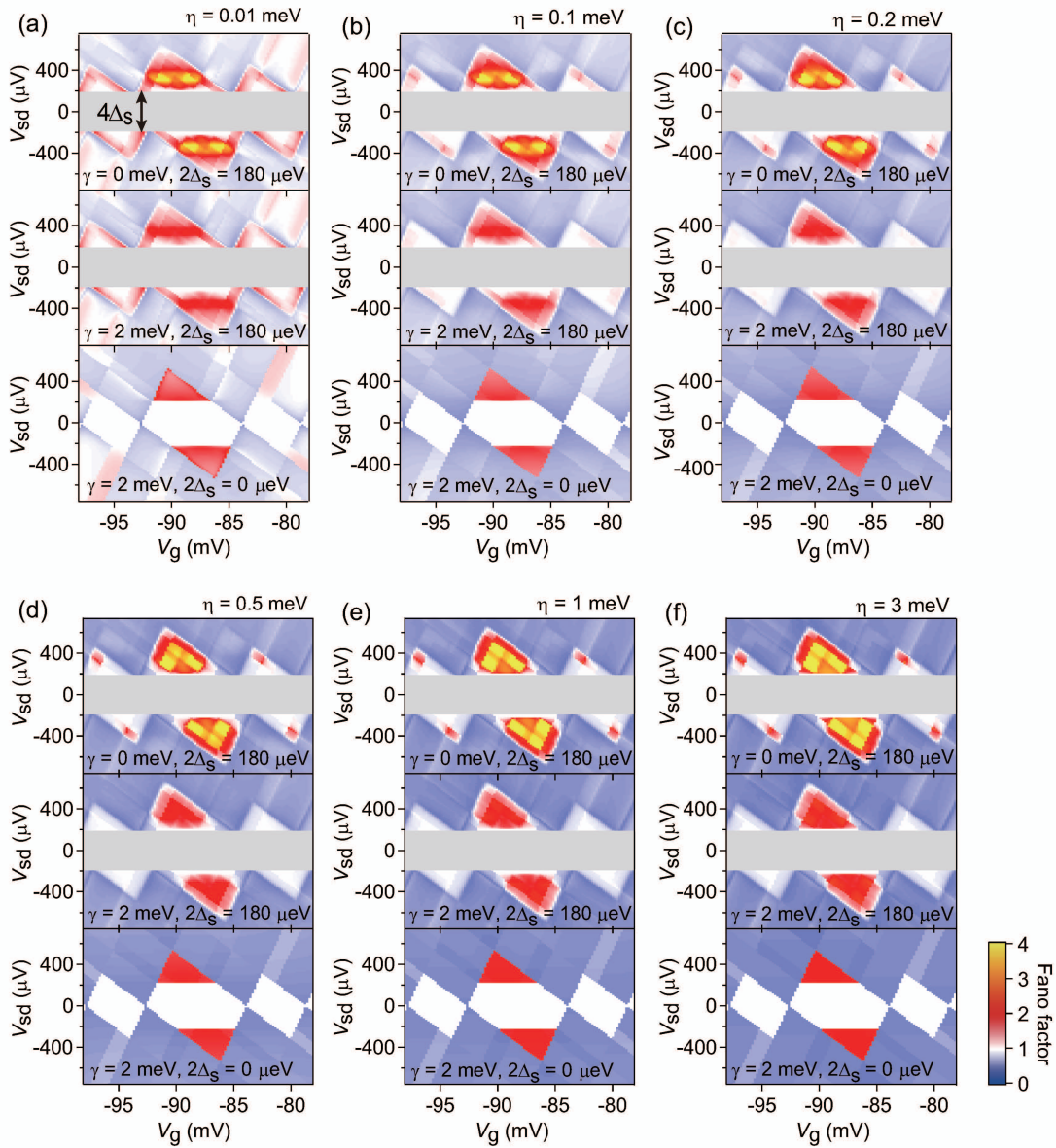

FIG. S2. Fano factor plotted as a function of $V_{sd}$ and $V_g$. The calculation is done with the parameters presented in the figures and texts. The center biggest Coulomb diamond corresponds to the even occupation. The details are described in the paper.

The figure set systematically compares the effects of the superconducting electrodes at zero field, the spin-flip relaxation, and decay of the intermediate states in cotunneling process. Each panel in Figs. (a) – (f) shows the results obtained with the parameters of (i) superconducting gap $2\Delta_s = 180$ $\mu$eV and spin-flip relaxation strength $\gamma = 0$ meV (top panel), (ii) $2\Delta_s = 180$ $\mu$eV and $\gamma = 2$ meV (middle panel), and (iii) $2\Delta_s = 0$ $\mu$eV and $\gamma = 2$ meV. Figures (a) – (f) show the dependence on the broadening $\eta$, i.e. the decay of the intermediate states in cotunneling.

From these simulations, we can see that various features in the data clarify the roles of the each effect. The comparison between top and middle panels highlights the role of the spin-flip relaxation, which reduces $F$ in the

Coulomb blockade regime. The middle and bottom panels show the effect of the superconducting gap. According to the presence of the superconducting gap, the Coulomb diamonds appear to be enlarged in the direction of $V_{\rm sd}$.

Moreover, by comparing with (a) – (f), we can see that the outlines of the Coulomb diamonds show $\eta$ dependence of $F$; $F$ is larger when $\eta$ is smaller. The $\eta$ dependence are also seen in the Coulomb blockade regimes, especially for even occupation (yellow area) and outside the Coulomb diamond (blue area).

### B. Parameters used to reproduce experiments

Here we summarize the details of the parameter values used for calculation shown in the paper (Table. I). As is clearly shown in the paper, the simulation based on these values reproduces our experiments. The definitions of the parameters are noted in the Section II. The value of $G_{r\pm}$ is estimated from the measured Coulomb peak width and we set $G_{r\pm} = 0.1$ meV commonly for simulation. The parameters used for Fig. S2 are the same values in Table I unless otherwise stated.

TABLE I. Summary of the parameters used to be compared with our experiments

|  | $G_{r\pm}$ (meV) | $U$ (meV) | $\Delta E$ (meV) | $2\Delta_s$ (meV) | $\Delta Z$ (meV) | $\alpha$ | $\beta$ | $V_{\rm g,0}$ (meV) |
|---|---|---|---|---|---|---|---|---|
| Data at 0 T | 0.1 | 0.227 | 0.309 | 0.180 | 0 | 0.237 | 0.0612 | -78.7 |
| Data at 0.9 T | 0.1 | 0.310 | 0.346 | 0 | 0.181 | 0.218 | 0.0683 | -89.1 |